\documentclass[aps,pre,floatfix]{revtex4}
\usepackage{epsf}
\usepackage{subfigure}
\usepackage{amsmath}
\usepackage{amssymb}
\usepackage{graphicx}
\usepackage{epstopdf}
\newcommand{\be}{\begin{equation}}
\newcommand{\ee}{\end{equation}}
\newcommand{\bea}{\begin{eqnarray}}
\newcommand{\eea}{\end{eqnarray}}
\newcommand{\diver}{\pmb{\nabla} \cdot}

\newcommand{\paren}[1]{\left( #1 \right)}
\newcommand{\mean}[1]{\left \langle #1 \right \rangle}

\begin{document}

(J. Stat. Mech. P02057, 2009)

\vspace{1cm}

\title{\bf Stochastic approach and fluctuation theorem for ion transport}

\author{David Andrieux and Pierre Gaspard}
\affiliation{Center for Nonlinear Phenomena and Complex Systems,\\
Universit\'e Libre de Bruxelles, Code Postal 231, Campus Plaine,
B-1050 Brussels, Belgium}
%(U.~L.~B.)

\begin{abstract}
We present a stochastic approach for ion transport at the mesoscopic
level. The description takes into account the self-consistent
electric field generated by the fixed and mobile charges as well as
the discrete nature of these latter. As an application we study the
noise in the ion transport process, including the effect of the
displacement current generated by the fluctuating electric field. The
fluctuation theorem is shown to hold for the electric current with
and without the displacement current.
\end{abstract}

%\pacs{82.20.Uv; 05.70.Ln; 02.50.Ey}

\maketitle

%%%%%%%%%%%%%%%%%%%%%%%%%%%%%%%%%%%%%%%%%%%%%%%%%%%%%%%%%%%%%%%%%%%%%%%%%%%%%%%%%%%%%%%%%%%

\section{Introduction}

The study of electrolyte solutions finds its origin in the seminal
works of Nernst \cite{N88,N89} and Planck \cite{P90a,P90b}. Ion
transport in electrolytes is described as arising from a
diffusive part due to the concentration gradients and a drift part
due to the external electric field imposed to the system. A more
accurate description is obtained if the electrical field is not only
generated by external means but also self-consistently incorporates
the contributions of the local deviations from electroneutrality
\cite{G10, C13}. Under the assumption that the electric field
propagates instantaneously - the quasi-static limit of the Maxwell
equations - this problem is known as the Nernst-Planck-Poisson
(NPP) problem \cite{B70,E00}. Here, the electric field
self-consistently arises from the distribution of charges as
described by the Poisson equation. In this case, the transport
process may present an additional contribution due to the temporal
variations of the electric field, known as the displacement current
\cite{J99,CC65,BB00}.
This approach has also played a basic role in the theory of other
systems: Indeed, the NPP equations, combined with statistical
arguments, lead for instance to the description of semiconductor p-n
junctions \cite{S49,R90}.

On the other hand, ionic solutions exist in many natural or artificial systems. 
Many biological processes crucially depend on the
transport of ions, e.g., between intra- and extracellular solutions \cite{A95,H01}.
Besides, solid-state nanopores are fabricated 
for the study of ionic current fluctuations \cite{SKDD08}.
In these systems, the transport of ion takes place in different geometries, 
either with the complexity of channel proteins
or more regular as straight channels in the case of nanopores.
Biological ion channels \cite{NE98}, as well as physico-chemical systems such
as nanopores \cite{RGCSM07}, nanofluidic diodes \cite{CS07}, 
or nanostructures studied by impedance spectroscopy \cite{BM05}, 
are often described by the NPP theory. 
Yet, they are of mesoscopic sizes and, at these
scales, the motion of ions in solution is subjected to molecular
fluctuations and presents a stochastic behavior. These random
fluctuations can be successfully described at the mesoscopic level as
Markovian stochastic processes \cite{NP71,S76,NP77,H05,AK07}. 
Non-Markovian discrete-state models have also been considered
to take into account the memory effects resulting, in particular,
from the lumping of diffusion \cite{BES96}. The description to be adopted
depends on the geometry. In short heterogeneous channels,
transport tends to proceed by jumps between discrete states corresponding to
wells in the free-energy potential. In long homogeneous channels
such as cylindrical nanopores, 
diffusion is not interrupted by barriers except at entrance and exit.
In any case, the master equation ruling these stochastic processes should reduce to the
evolution equation for the charges densities at the macroscopic level.
In this respect, one of the fundamental problems is to incorporate
the long-range Coulomb interaction between the electric charges
in the description.

In the present work, our purpose is to present a description of ion transport 
which is consistent with the laws of both electricity and statistical thermodynamics. 
The key point is that the long-range Coulomb interaction 
deeply influences the fluctuations of the particle and total currents. 
The proposed model describes the spatial distribution of the discrete
number of ions in a long channel. 
The ions undergo random jumps due to
the thermal agitation in a self-consistently generated electrical
potential landscape. This description takes into account the
stochastic aspects of the time evolution as well as the
self-consistent electric field generated by the charge distribution.
In this regard, it offers a computationally favorable alternative to
more expensive techniques such as Brownian dynamics \cite{TLF77,KAC01}.
This approach allows us to study various aspects of ion transport, 
revealing, e.g., the long-ranged spatial correlations
of the potential fluctuations inside the channel.
We further study the fluctuations present in the transport process,
including the contributions of the displacement current.
Also, we show that the fluctuation theorem \cite{ECM93, GC95, LS99,
M99, S05, AG07JSP, HS07, SPWS08} describing the large-deviation 
properties of the current holds in this case as well.

   The paper is organized as follows.  In Sec. \ref{macro}, we
introduce the macroscopic description of ion transport in
terms of the Nernst-Planck-Poisson equations.
   The stochastic description is presented in Sec. \ref{stoch}. The
fluctuations in the ion transport process are studied in Sec.
\ref{noise}.  In Sec.  \ref{sec.ft}, the symmetry of the fluctuation
theorem is shown to hold for both the particle and electric currents.
Conclusions are drawn in Sec. \ref{conclusions}.

\section{Macroscopic equations}
\label{macro}

We consider a one-dimensional conduction channel extending from $x=0$ to $x=\ell$.
An important class of electrolytic systems are those which contain a
homogeneous distribution of immobile ions \cite{T53}.
We therefore consider a model with a single mobile species of
charge $e$ and density $n$ being transported in a channel presenting
a fixed ion density $n_-$ of opposite charge $-e$.
The charge density inside the system is thus given by
\bea
\rho = e(n-n_-) \, .
\eea
The case of several fixed and mobile species can be treated in a
similar way. Expressing the electric field $\pmb{\mathcal E}$ in
terms of the electric potential $\Phi$ as $\pmb{\mathcal E} = -
\pmb{\nabla} \Phi$,
the problem is ruled by the coupled diffusion and Poisson equations:
\begin{subequations} \label{coupled}
\bea
\label{dn}
\partial_t n &=& \pmb{\nabla} \cdot \paren{ D\pmb{\nabla} n + \mu e n
\pmb{\nabla} \Phi} \, , \\
\label{poisson}
\nabla^2 \Phi &=& -\frac{e}{\epsilon}(n-n_-) \, .
\eea
\end{subequations}
$D$ is the diffusion coefficient of the mobile species and $\mu$ is
the mobility coefficient which is given by Einstein's relation $\mu =
D/k_{{\rm B}}T$.
The dielectric fluid is here homogeneous throughout the channel and
has a dielectric constant of $\epsilon$.
Equation \eqref{dn} is the conservation equation for the particle
density, $\partial_t n = - \diver \pmb{J}$, where $\pmb{J}$ is the
particle current density. The Poisson equation \eqref{poisson} in
turn gives the electric potential as a function of the charge
density. This coupled system \eqref{coupled} forms the
Nernst-Planck-Poisson equations. They correspond to the quasi-static
limit of the complete Maxwell equations.
On the other hand, the total current density reads \cite{J99,BB00}
\bea
\pmb{I} = e \pmb{J} +\epsilon \frac{\partial \pmb{\mathcal E}}{\partial t} \, ,
\label{ijd}
\eea
the last term being the displacement current density arising from
temporal variations of the electrical field. The displacement current
vanishes in a stationary state but contributes, for example, when the
system relaxes toward the stationary state. Note that, by virtue of
the Poisson equation \eqref{poisson}, the total current density is
divergence free, $\pmb{\nabla} \cdot \pmb{I}=0$, {\it at all
times}.
We further notice that the experimentally measurable quantity is
given by the total current \cite{BB00}.

The channel is in contact with two reservoirs of particles maintained
at fixed concentrations and electric potentials. Equations \eqref{coupled} are thus
supplemented by the boundary conditions
\begin{subequations} \label{bound.2}
\bea
\label{bound.n}
n(0)&=& n_{\rm L} \, , \quad n(\ell) = n_{\rm R} \, ,\\
\label{bound.phi}
\Phi(0)&=& \Phi_{\rm L}\, , \quad \Phi(\ell) = \Phi_{\rm R} \, ,
\eea
\end{subequations}
determining the concentrations on the left (L, $x=0$) and right (R,
$x=\ell$) boundaries as well as the external potential difference
$V=\Phi_{\rm L}-\Phi_{\rm R}$ applied to the channel.

For time-independent boundary conditions, the system evolves toward a
stationary state where $\partial_t n =0$ so that the particle current
density $\pmb{J}$ is constant in time and in space. For the
one-dimensional channel here considered, the stationary density
satisfying the boundary conditions \eqref{bound.2}
is expressed as
\bea
n(x)= n_{\rm L} {\rm e}^{\phi(x)-\phi_{\rm L}} - \frac{J}{D}\,  {\rm
e}^{\phi(x)} \int_0^x {\rm e}^{-\phi(y)}\, dy \, ,
\eea
along with the particle current
\bea
J=-D\, \frac{n_{\rm R} {\rm e}^{-\phi_{\rm R}} - n_{\rm L} {\rm
e}^{-\phi_{\rm L}} }{\int_0^\ell {\rm e}^{-\phi(x)}dx } \, ,
\label{J}
\eea
where we introduced the dimensionless potential $\phi\equiv
-e\Phi/k_{{\rm B}}T$ to simplify the expressions. This potential
obeys the dimensionless Poisson equation
\be
\label{adim-poisson}
\nabla^2 \phi = \frac{1}{\lambda^2}\left (\frac{n}{n_-}-1\right) \, ,
\ee
where
\be
\lambda \equiv \sqrt{\frac{\epsilon k_{{\rm B}}T}{e^2 n_-}}
\label{debye}
\ee
is the Debye screening length of the system \cite{DH23}.

The stationary state is a state of thermodynamic equilibrium if the
particle current vanishes, i.e., when we have
\bea
(\phi_{\rm L} - \phi_{\rm R})_{{\rm eq}} = \ln \frac{n_{\rm 
L}}{n_{\rm R}} \qquad {\rm or} \qquad (\Phi_{\rm L} - \Phi_{\rm 
R})_{{\rm eq}} = \frac{k_{{\rm B}}T}{e} \ln \frac{n_{\rm R}}{n_{\rm 
L}} \, .
\label{eq}
\eea
This voltage difference is known as the Nernst potential
\cite{N88,N89}. If this condition is not satisfied we are in a
nonequilibrium situation characterized by the presence of a
non-vanishing ionic current given by Eq. \eqref{J}. In the stationary
state, the associated total current reads
\bea
I = eJ
\eea
since the displacement current vanishes in this case. However, at the
mesoscopic level, fluctuations in the charge distribution generate a
fluctuating displacement current, as studied in Sec. \ref{noise}.

\section{Stochastic description}
\label{stoch}

The study of noise and molecular fluctuations at the mesoscopic scale
is successfully accomplished in terms of a Markovian random
processes. Indeed, the master equation is known to describe the
fluctuations down to the mesoscopic scale \cite{NP71,S76,NP77}.

We here introduce a stochastic model for the distribution of ions in
the channel that incorporates the self-consistent electric field
generated by the fluctuating distribution of ions.
The channel is divided into $L$ cells of volume $\Delta V$, cross
section $\sigma$, and length $\Delta x = \ell/L$ centered at the
positions $x_i= (i-1/2)/\Delta x$ ($i=1, \dots,L$).
Each cell contains a discrete number of mobile ions $N_i$ and a
constant number $N_- = n_- \Delta V$ of fixed ions. The cells $0$ and
$L+1$ correspond to the external reservoirs maintained at fixed
concentrations so that their particle numbers $N_0$ and $N_{L+1}$
remain constant in time.
The dimensionless electric potential $\phi_i =- e\Phi_i / k_{{\rm
B}}T$ is defined on each cell as well. It obeys the Poisson equation \eqref{adim-poisson} which is discretized according to
\bea
\frac{\phi_{i+1}-2 \phi_i+\phi_{i-1}}{\Delta x^2} =
\frac{1}{\lambda^2} \paren{\frac{N_i}{N_-}-1}
\label{disc.poisson}
\eea
with the boundary conditions $\phi_0=\phi_{\rm L}$ and
$\phi_{L+1}=\phi_{\rm R}$.
Equations \eqref{disc.poisson} form a
linear system that must be solved at each time the particle
distribution changes. Its exact solution is given in Appendix
\ref{appA}. The extension of the model to several (positive or
negative) ion species and inhomogeneous distribution of fixed ions
and permittivity is straightforward.

The state of the system is determined by the number of ions $N_i$ in
each cell. The evolution equation for the probability distribution
$P(N_1,...,N_L)$ to observe a configuration $\{ N_i\}$ of particles
is ruled by the master equation
\bea
\frac{d}{dt}P(N_1,...,N_L) = \sum_{i=0}^{L} &\Big[&
W_{+i}(...,N_i+1,N_{i+1}-1, ...)P(...,N_i+1,N_{i+1}-1, ...)\nonumber
\\
&-& W_{+i}(...,N_i,N_{i+1}, ...)P(...,N_i,N_{i+1}, ...) \nonumber  \\
&+& W_{-i}(...,N_i-1,N_{i+1}+1, ...)P(...,N_i-1,N_{i+1}+1, ...) \nonumber \\
&-& W_{-i}(...,N_i,N_{i+1}, ...)P(...,N_i,N_{i+1}, ...) \ \Big]
\label{ME}
\eea
where $W_{\pm i}(\cdot)$ denotes the transition rate between two
configurations of the system. The transition $\pm i$ changes the
configuration from $(...,N_i,N_{i+1}, ...)$
to $(...,N_i \mp 1
,N_{i+1}\pm 1,...)$.
These transitions rates are supplemented with the boundary conditions
that $N_0$ and $N_{L+1}$ take fixed values.

The transition rates can be expressed as
\bea
W_{+i}(...,N_i,N_{i+1}, ...) &=& \psi(\Delta U_{i,i+1}) N_i \\
W_{-i}(...,N_i,N_{i+1}, ...) &=& \psi(\Delta U_{i+1,i}) N_{i+1}
\label{W}
\eea
with the function
\bea
\psi(\Delta U) = \frac{D}{\Delta x^2} \frac{\beta\Delta U}{{\rm
e}^{\beta\Delta U}-1}
\label{dfn_psi}
\eea
where $D$ is the diffusion coefficient and $\beta= (k_{{\rm
B}}T)^{-1}$ the inverse temperature. The prefactor $D/\Delta x^2$
allows us to recover the appropriate evolution equation \eqref{dn} in
the macroscopic limit, $L\rightarrow \infty$ \cite{G90}.
The transition rates are proportional to the number of mobile ions in
the cells and depend on the free electrostatic energy difference
$\Delta U_{i,i+1}$ as a result of the transition event:
\bea
\Delta U_{i,i+1} = \frac{e}{2} (V_i + V'_i) \, .
\label{dE}
\eea
In Eq. \eqref{dE}, $V_i$ and $V'_i$ are the voltage drops across the
cells $i$ and $i+1$ before and after a transition event,
respectively. This expression corresponds to the change in
electrostatic energy resulting from the transition event, as shown in
Appendix \ref{appA}.
Consequently, Eq. \eqref{dE} can be expressed as
\bea
\Delta U_{i,i+1} = \frac{e}{2} (\Phi_{i+1}-\Phi_i+\Phi'_{i+1}-\Phi'_i)
\label{dEphi}
\eea
in terms of the electric potential calculated for the current
configuration, $\Phi_j = \Phi_j(N_1,...,N_L)$, and the
electric potential $\Phi'_j$ that would occur if the corresponding transition
$\pm i$ were performed:
\bea
\Phi'_j = \Phi'_j(...,N_i \mp 1 ,N_{i+1}\pm 1, ...) \, .
\eea
The potential $\Phi'_j$ can be expressed as
\bea
\Phi'_j = \Phi_j \mp e \paren{{\pmb{\mathsf C}}^{-1}}_{j,i} \pm e
\paren{{\pmb{\mathsf C}}^{-1}}_{j,i+1}
\label{phi'}
\eea
in terms of the matrix ${\pmb{\mathsf C}}^{-1}$ obtained from the
solution of the discretized Poisson equation (see Appendix
\ref{appA}).

The function $\psi$ satisfies the identity
\bea
\psi(\Delta U) = \psi(-\Delta U) {\rm e}^{-\beta \Delta U}
\label{det_bal}
\eea
which guarantees that detailed balance is fulfilled at equilibrium
%PG
\cite{O31,T63,footnote}.
On the other hand, the nonequilibrium constraints or affinities
driving the system out of equilibrium \cite{DD36} can be identified
in the stochastic description thanks to a construction put forward by
Schnakenberg \cite{S76}, according to which the affinities are
obtained from the cyclic trajectories in the forward and backward
directions. In our case, as verified in Appendix \ref{appB}, the
macroscopic affinity is readily identified by considering
trajectories involving the transfer of an ion from one reservoir to
the other, yielding
\bea
A = \ln \paren{\frac{N_0}{N_{L+1}}{\rm e}^{\Delta\phi}} = \ln
\paren{\frac{n_{\rm L}}{n_{\rm R}}{\rm e}^{\Delta\phi}} \, .
\label{aff}
\eea
Here, $\Delta \phi = \phi_{\rm R}-\phi_{\rm L}$ and we note that this
affinity only involves macroscopic quantities as it should. At
equilibrium the affinity vanishes, $A_{{\rm eq}}=0$, and we recover
the macroscopic equilibrium condition \eqref{eq}.

The stochastic process is simulated as follows. Random trajectories of the
system are obtained using Gillespie's algorithm \cite{G76}, which is
known to reproduce the statistical properties of the master equation (\ref{ME}).
At each random jump, we have to recalculate the electric potential in
the channel by solving the system \eqref{disc.poisson}.
The simulations are performed with the following parameters. Each cell
contains a number $N_-=25$ of fixed ions. By an appropriate choice of
the time unit, we can impose for example $D/\Delta x^2 = 1$. Using
the dimensionless electric potential $\phi$, the only remaining
dimensional parameter is the Debye length $\lambda$. We form the
dimensionless quantity $\ell/\lambda$, i.e., the ratio between the
channel and Debye's length, for which we choose the value
$\ell/\lambda=50$. For a Debye length of $7 \, {\rm nm}$ this
would correspond to a channel length of $350 \, {\rm nm}$.

\begin{figure}[t]
\begin{tabular}{cc}
%\vspace*{0.8cm}
\rotatebox{0}{\scalebox{0.33}{\includegraphics{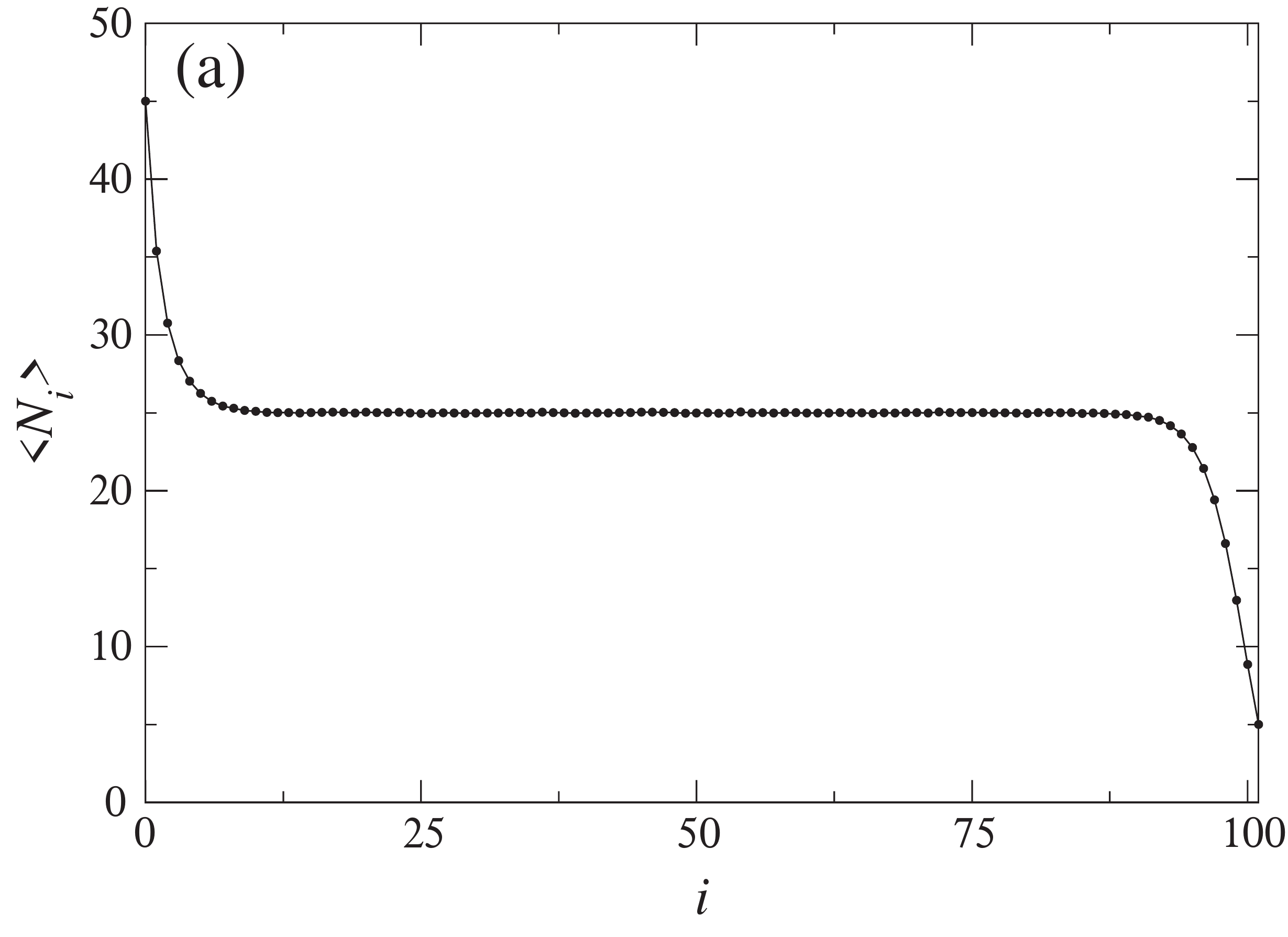}}} &
\rotatebox{0}{\scalebox{0.33}{\includegraphics{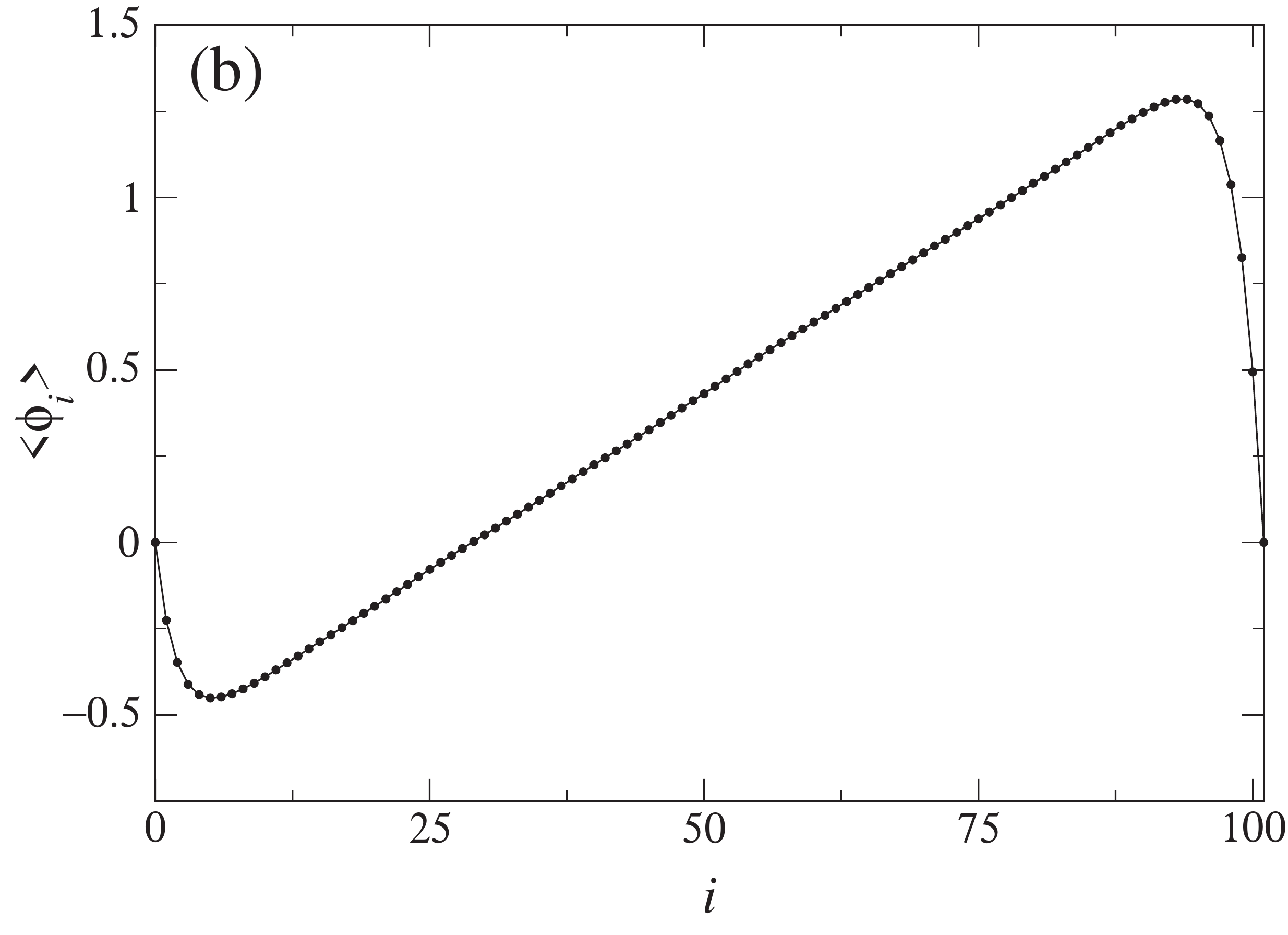}}} \\
\end{tabular}
   \caption{(a) Mean number of ions $\mean{N_i}$ as a function of the
position $i$ in the channel. (b) Mean dimensionless electric
potential $\mean{\phi_i}$ as a function of the position $i$ in the
channel. The channel is composed of $L=100$ cells, 
each with $N_- = 25$ fixed ions.  The channel is submitted to
the boundary conditions $N_{\rm L} = 45$, $N_{\rm R} = 5$, and
$\phi_{\rm L}=\phi_{\rm R}$.}
\label{fig1}
\end{figure}

The link with the macroscopic description is established in the limit
where the volume of the cells $\Delta V$ vanishes in which case the
concentrations are recovered as $n(x_i)=\mean{N_i}/\Delta V$. In Fig.
\ref{fig1}a, we depict the average number of ions along the channel,
estimated by the time average
\bea
\mean{N_i} = \lim_{T\rightarrow \infty} \frac{1}{T} \int_0^T N_i(t) dt
\eea
which, by ergodicity, is equivalent to the ensemble average
$\mean{N_i} = \sum N_i P(..., N_i, ...)$.
We see that electroneutrality, $\mean{N_i} \simeq N_-$, is well
respected except for a small layer from the boundaries over a
distance of the order of the screening distance $\lambda$. The
corresponding mean electric potential $\mean{\phi_i}$
is shown in Fig. \ref{fig1}b. It is approximately linear throughout
the channel, which corresponds to a constant electric field, except
again for a small layer near the boundaries.

\begin{figure}[t]
\begin{tabular}{cc}
%\vspace*{0.8cm}
\rotatebox{0}{\scalebox{0.33}{\includegraphics{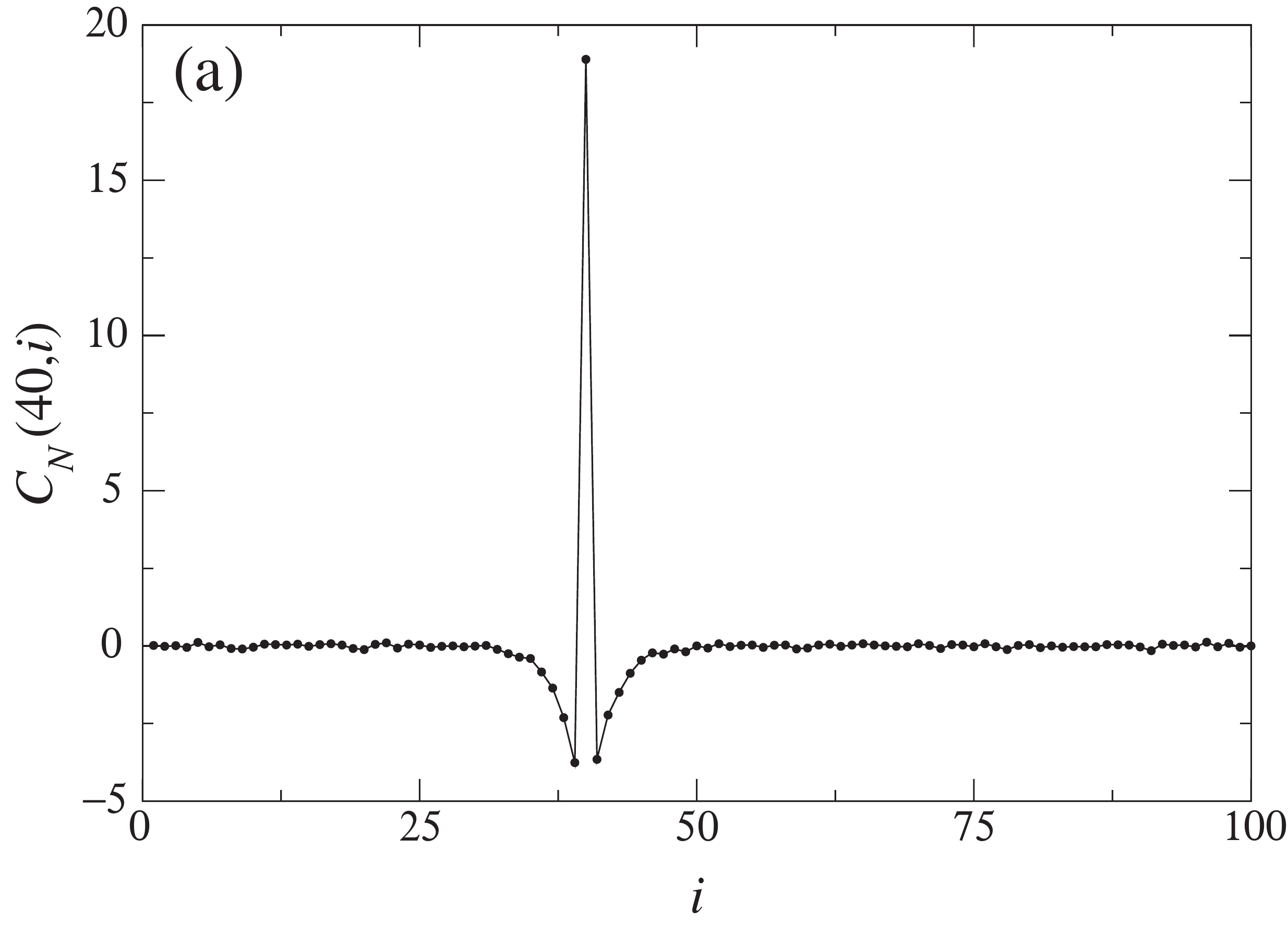}}} &
\rotatebox{0}{\scalebox{0.33}{\includegraphics{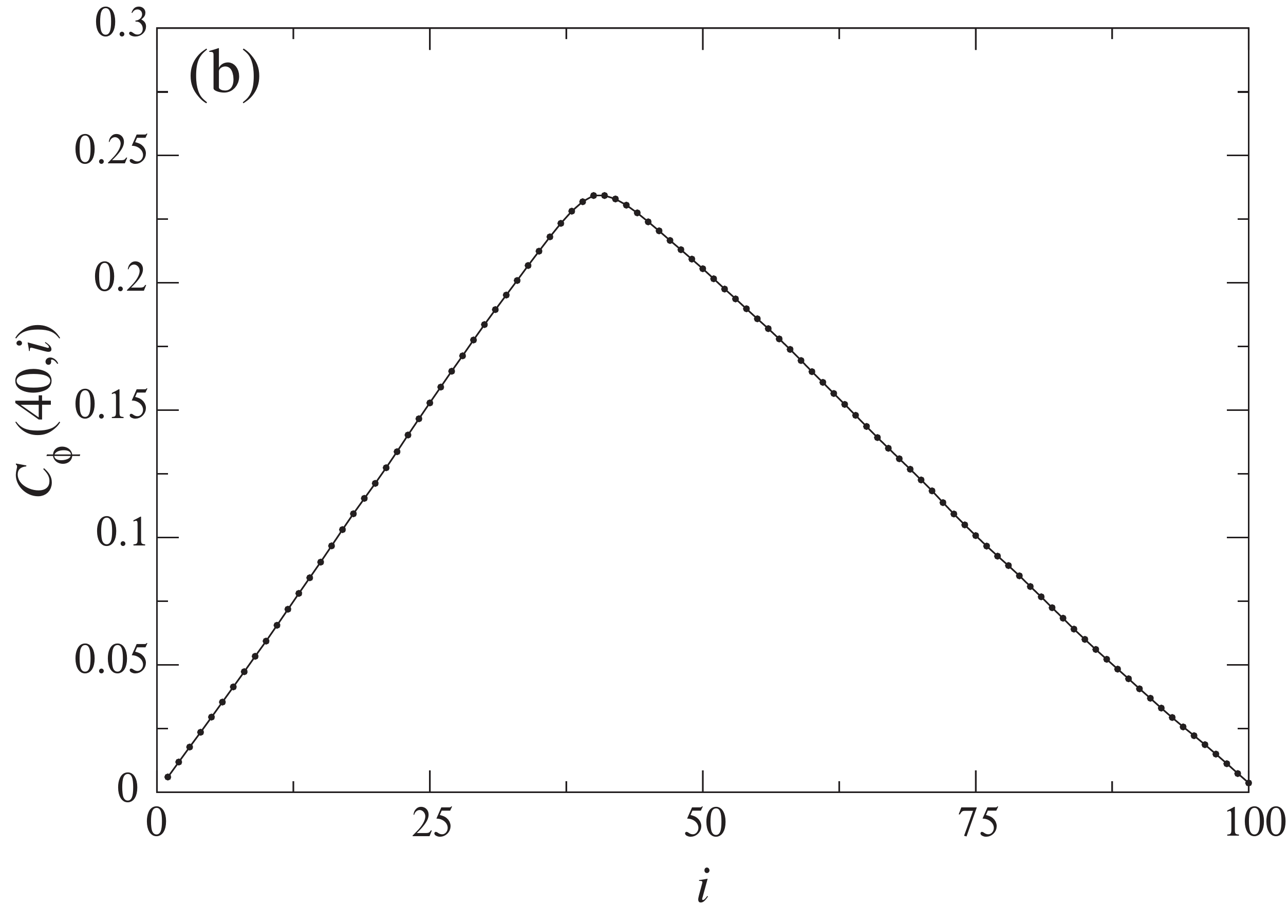}}} \\
\end{tabular}
   \caption{(a) Spatial correlations $C_N (40,i)$ for the density
fluctuations between site $40$ and site $i$ as a function of the position $i$ in the
channel.
(b) Spatial correlations $C_\phi (40,i)$ for the electric potential
fluctuations between site $40$ and site $i$ as a function of the position $i$ in the
channel. The channel is composed of $L=100$ cells, each containing $N_- = 25$ fixed ions.
The boundary conditions are $N_{\rm L} = 45$, $N_{\rm R} = 5$, and $\phi_{\rm
L}=\phi_{\rm R}$.}
\label{fig2}
\end{figure}

Due to the electric interaction, the particle density fluctuations
present spatial correlations along the channel. The correlations
between cells are obtained by the time averages
\bea
C_N(i,j) \equiv \mean{N_i N_j} - \mean{N_i}\mean{N_j} =
\lim_{T\rightarrow \infty} \frac{1}{T} \int_0^T \left[
N_i(t)-\mean{N_i}\right] \left[N_j(t)-\mean{N_j} \right] dt \, .
\eea
Fig. \ref{fig2}a shows the effect of the electric interaction on the
spatial correlations. We observe negative correlations between
neighboring sites, showing that an excess charge at some location
induces a repulsive effect on the neighboring sites. The negative
correlations rapidly diminish with the separation between sites over
a distance comparable to the Debye length. The presence of such
correlations also shows that the particle distribution is not
Poissonian even at equilibrium in contrast with population dynamics
of chemical reactions network \cite{NP77}.
The electric potential correlation function
\bea
C_\phi(i,j) \equiv \mean{\phi_i \phi_j} - \mean{\phi_i}\mean{\phi_j}
\eea
is depicted in Fig. \ref{fig2}b. It presents long-range correlations
with a linear decrease in both directions between the site and the
boundaries.

\section{Fluctuations in ion transport}
\label{noise}

In this section, we study the fluctuations in the transport process.
For this purpose, we first introduce the fluctuating particle current
$j_k$ counting the number of ions transferred between the cells $k$
and $k+1$. We denote by $\varepsilon_k(s) = \pm 1$ the discrete
charge transfer of an ion in the positive or negative direction
between the cells $k$ and $k+1$ during the $s^{{\rm th}}$ random
transition occurring at the time $t_s$ [$\varepsilon_k(s) =0$
otherwise]. Accordingly, the fluctuating particle current
is given
by
\bea
j_{k}(t) = \sum_{s=-\infty}^{+\infty} \varepsilon_k(s) \delta (t-t_s) \, .
\label{ji}
\eea
We notice that the currents \eqref{ji} are so-called point processes
composed of singular events occurring at random times.
In the stationary state, the average particle current crossing the
channel is time-independent and can be obtained from the stationary
probability distribution as
\bea
J \equiv \mean{j_k(t)} = \sum_{N_1,...,N_L} P_{{\rm st}}
(N_1,...,N_L) [W_{+k}(N_1,...,N_L) - W_{-k}(N_1,...,N_L)] \, .
\label{meanJ}
\eea
By current conservation, the average particle current
is independent of the position in the channel, $J \equiv \mean{j_k}$
for all $k$.

However, the experimentally measured electric current is composed of
the particle and displacement currents according to Eq. \eqref{ijd}.
Indeed, even in a stationary state, fluctuations in the particle
distribution at the mesoscopic level generate a fluctuating electric
field that, in turn, generates a fluctuating contribution to the
electric current. We thus calculate the change in the electric field
associated with the random jumps of the charge carriers. The change
in the electric field between the cells $k$ and $k+1$, $E_{k+1/2} =
-(\Phi_{k+1}-\Phi_k)/\Delta x$, is expressed as
\bea
-\frac{1}{\Delta x} \Big( \Phi'_{k+1} - \Phi'_{k} -\Phi_{k+1} +
\Phi_{k} \Big) = \pm \frac{e}{\Delta x} \Big[\paren{{\pmb{\mathsf
C}}^{-1}}_{k+1,l} - \paren{{\pmb{\mathsf C}}^{-1}}_{k+1,l+1} -
\paren{{\pmb{\mathsf C}}^{-1}}_{k,l} +  \paren{{\pmb{\mathsf
C}}^{-1}}_{k,l+1}\Big]
\eea
if the transition $W_{\pm l}$ is performed. The second equality is
obtained by using Eq. \eqref{phi'} for the potential $\Phi'(...,N_l
\mp 1 ,N_{l+1}\pm 1, ...)$ after the transition.
The associated displacement current is given by $\epsilon\sigma$
times the change in the electric field,
where $\sigma$ is the cross section of the channel ($\sigma\Delta x=
\Delta V$).
The displacement current associated with the transition $W_{\pm
l}$ and measured between the cells $k$ and $k+1$ thus reads
\bea
\label{dk}
\begin{cases}
                   \pm e\frac{1}{L+1} & \text{if} \quad k\not= l \\
                   \mp e\frac{L}{L+1}& \text{if} \quad k=l
                  \end{cases}
\eea
where we used expression \eqref{c-1} and the equality $\epsilon \sigma = \alpha\Delta x$. Now, the
contribution of the transition $W_{\pm l}$ to the total current is
the sum of the particle current, $\pm e$ if $k=l$ and zero otherwise,
and the displacement current \eqref{dk} so that
the location $l$
contributes to the total current by the amount
\bea
\frac{e}{L+1}\, j_l(t)
\eea
as it should since the electric current must be divergence free at
all times. Remarkably, all the transitions $\pm l$ contribute equally
to the electric current.
The total current is finally obtained by summing the contributions
from all the transitions $\pm l$:
\bea
I(t) = \frac{e}{L+1} \sum_{l=0}^{L} j_l (t) \, .
\label{Itot}
\eea
The sum runs over all the possible charge transfers since each charge
displacement in the system modifies the electric field, inducing in
turn a displacement current.
Expression \eqref{Itot} can also be obtained from the Ramo-Shockley
theorem \cite{BB00,R39,S38,NPGE04} linking the current flowing in the
external circuit to charge movement inside the system. By virtue of Eq.
\eqref{meanJ}, the mean total current is related to the mean particle
current according to
\bea
\mean{I} = e J \, ,
\eea
showing that the displacement current does not contribute on average,
as it should. However, as shown below, the particle and total
currents differ in the properties of their non-zero frequency
fluctuations.

\begin{figure}[t]
\centerline{\scalebox{0.33}{\includegraphics{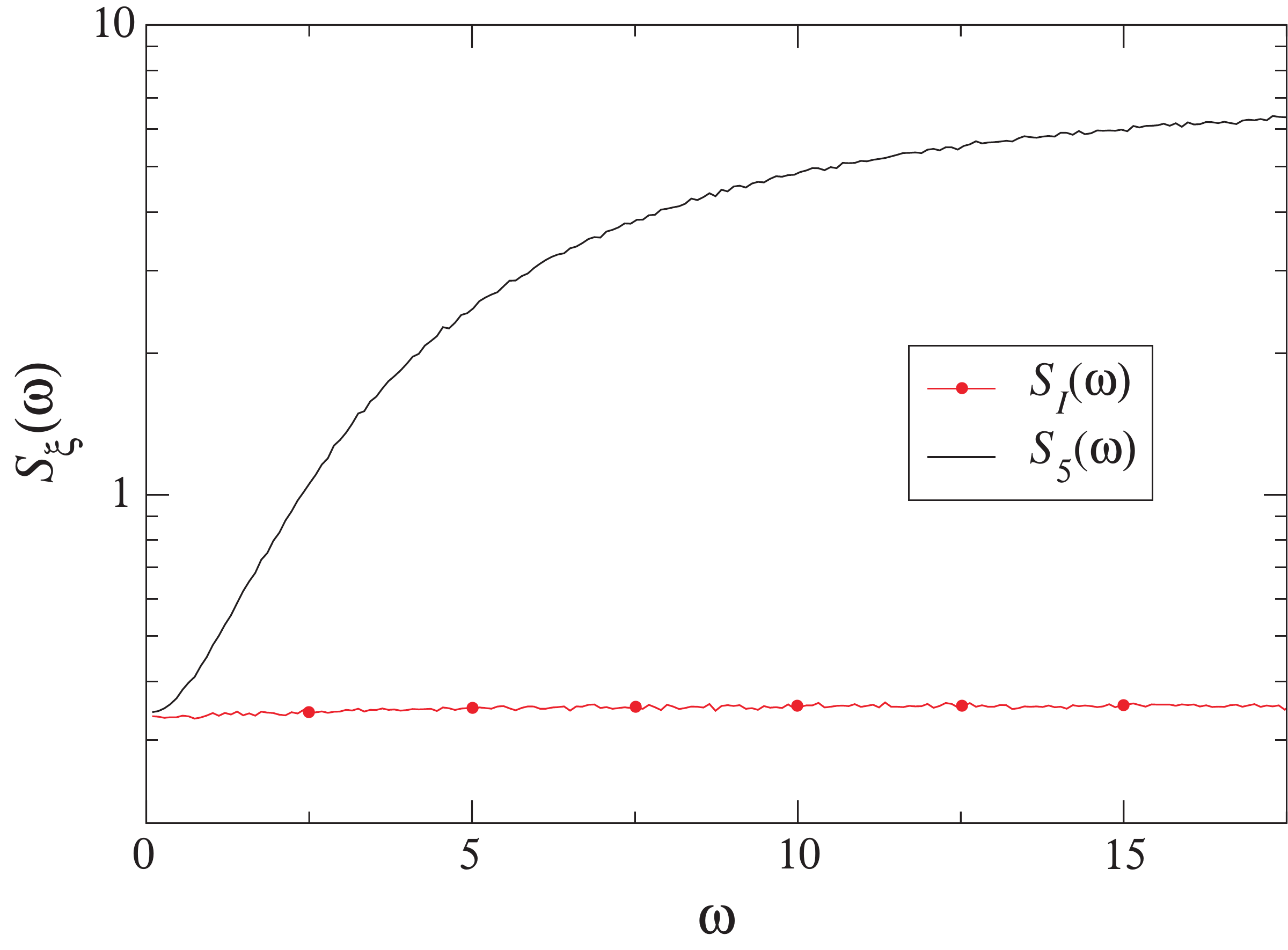}}}
\caption{Power spectra of the particle current (evaluated at $k=5$)
and of the total current in units of $e^2$ (line with dots).
The channel is composed of $L=20$ cells and is submitted to the
boundary conditions $N_{\rm L} = 5$, $N_{\rm R} = 45$, and $\phi_{\rm
L}=\phi_{\rm R}$. Each cell contains $N_- = 25$ fixed ions.}
\label{fig3}
\end{figure}

Besides the mean currents, the noise characterization offers further
information on the transport process \cite{BB00}. The power spectra
of the currents are defined as the Fourier transform of the current
correlation function:
\bea
S_{\xi}(\omega)= \frac{1}{2\pi} \int_{-\infty}^{+\infty} {\rm
e}^{i\omega t} \Big[ \mean{\xi(t)\xi(0)} -\mean{\xi}^2 \Big] dt \, ,
\eea
where $\xi$ is associated with either the particle current
correlations, $\xi =j_k$, or to the electric current, $\xi=I$.
These power spectra can be obtained from an ensemble of long random
trajectories according to
\bea
S_\xi(\omega)= \lim_{T\rightarrow \infty} \frac{1}{2\pi T}
\mean{\vert \tilde{\xi}(\omega)\vert^2} - \mean{\xi}^2 \delta(\omega)
\eea
where $\tilde{\xi}(\omega) \equiv \int_0^T {\rm e}^{-i\omega t}
\xi(t)dt$ and where $\mean{\cdot}$ denotes an average over several
trajectories. For the particle current, we thus have
\bea
\tilde{j}_{k}(\omega)= \int_0^T {\rm e}^{-i\omega t} j_k(t) dt =
\sum_{s} \varepsilon_k(s) \, {\rm e}^{-i\omega t_s} \, .
\eea
Similarly, the power spectrum of the total current \eqref{Itot} is
obtained from
\bea
\tilde{I} (\omega) = \int_0^T {\rm e}^{-i\omega t} I(t) dt =
\frac{e}{L+1} \sum_{l} \sum_{s} \varepsilon_l(s) \, {\rm e}^{-i\omega
t_s}
\eea
where the sums take into account every transition in the system, in
accord with Eq. \eqref{Itot}.

In Fig. \ref{fig3}, we depict the power spectrum of the particle and
total currents.
We see that, under the same conditions, they are qualitatively
different from each other, showing that the inclusion of the
displacement current affects the fluctuations at positive
frequencies and therefore must be properly taken into account. The 
power spectrum of the particle current increases with the frequency 
$\omega$ because the corresponding correlation function is negative. 
Such anticorrelations constitute another signature of the Coulomb 
repulsion between the ions, which reduces the probability of a 
definite transition event after jumping through the same section in 
the same direction. This frequency dependence of the power spectrum
reflects the memory effects induced by the long-range Coulomb interaction
on the random motion of the particles.
On the other hand, the total current presents a strongly reduced noise at
all frequencies because of the long-range correlations induced by the
changes in the distribution of charges.
We note that the power spectra reach a constant positive value in the
high-frequency limit that results from the intrinsic randomness found
at all times scales in such stochastic processes \cite{GW93}.
Precisely, in this limit, the spectra $S_k$ are given by the
shot-noise formula
\bea
S_k (\infty) = \sum_{N_1,...,N_L} P_{{\rm st}} (N_1,...,N_L)
[W_{+k}(N_1,...,N_L) + W_{-k}(N_1,...,N_L)] \, .
\label{Sk}
\eea
In addition, in the high-frequency limit, the different noise
processes become uncorrelated so that
\bea
S_I (\infty) = \frac{e^2}{(L+1)^2} \sum_{k=0}^L S_k (\infty) \, .
\label{SI}
\eea
In the bulk of the channel away from the boundaries by a few
Debye's lengths, the spectra $S_k(\omega)$ turn out to be uniform,
i.e., approximately independent of the location $k$.
This explains
that, for a channel longer than Debye's length, we approximately find
\bea
S_I (\infty) \simeq \frac{e^2}{L+1} S_k (\infty) \, ,
\eea
as observed in Fig. \ref{fig3}.

Furthermore, when the applied voltage $V=\Phi_L - \Phi_R$ is 
sufficiently large we may neglect the backward transitions, 
$W_{-k}(N_1,...,N_L) \simeq 0$, in which case Eqs. \eqref{meanJ} and 
\eqref{Sk} coincide so that
\bea
S_k (\infty) = \frac{\mean{I}}{e}
\eea
and
\bea
S_I (\infty) = \frac{e^2}{L+1} S_k (\infty) = \frac{e \mean{I}}{L+1}
\eea
by virtue of Eq. \eqref{SI}. These expressions correspond to the 
Schottky noise formulae \cite{S18}.

\section{Fluctuation theorem for the currents}
\label{sec.ft}

\begin{figure}[t]
\begin{tabular}{cc}
%\vspace*{0.8cm}
\rotatebox{0}{\scalebox{0.33}{\includegraphics{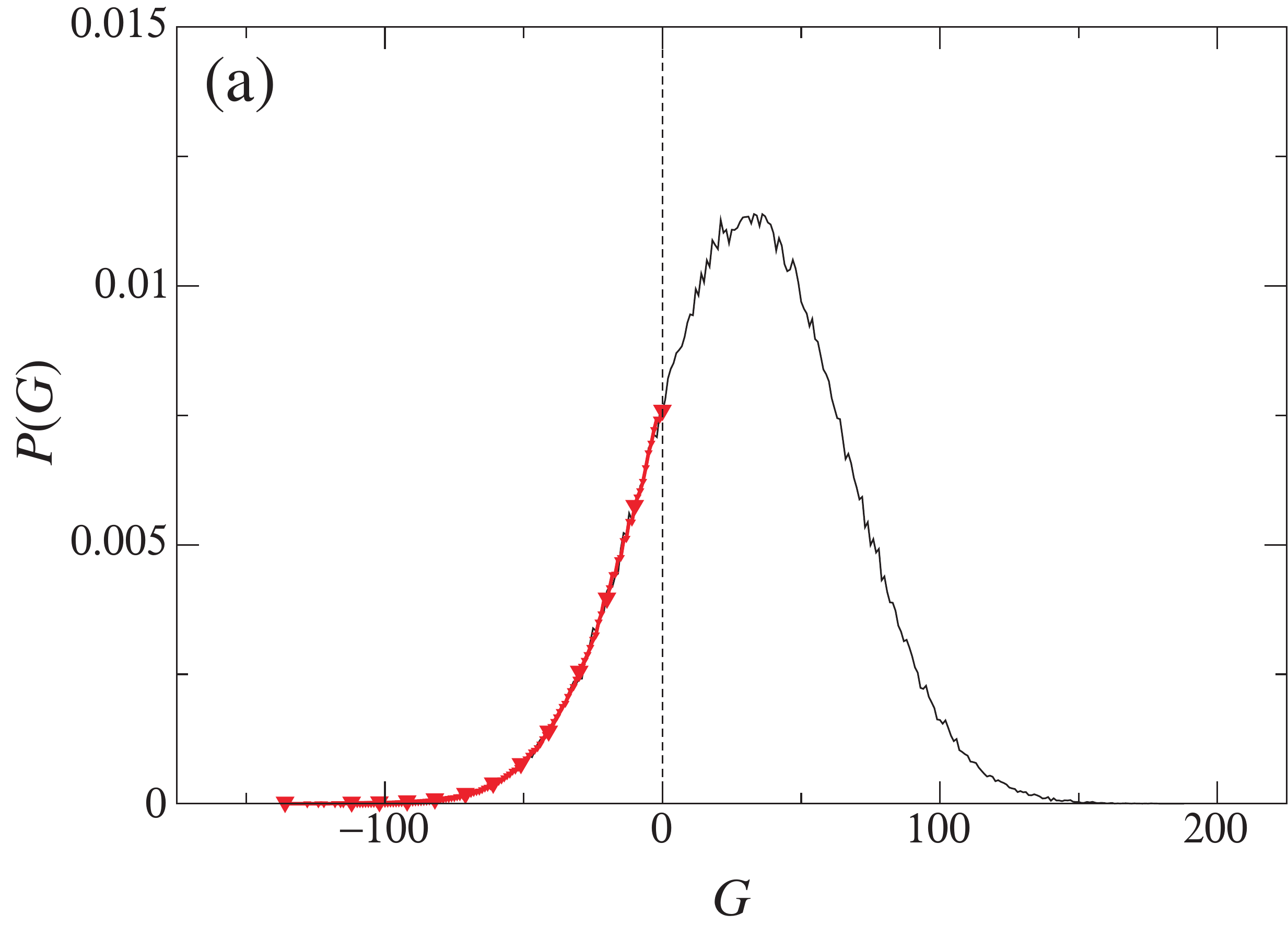}}} &
\rotatebox{0}{\scalebox{0.33}{\includegraphics{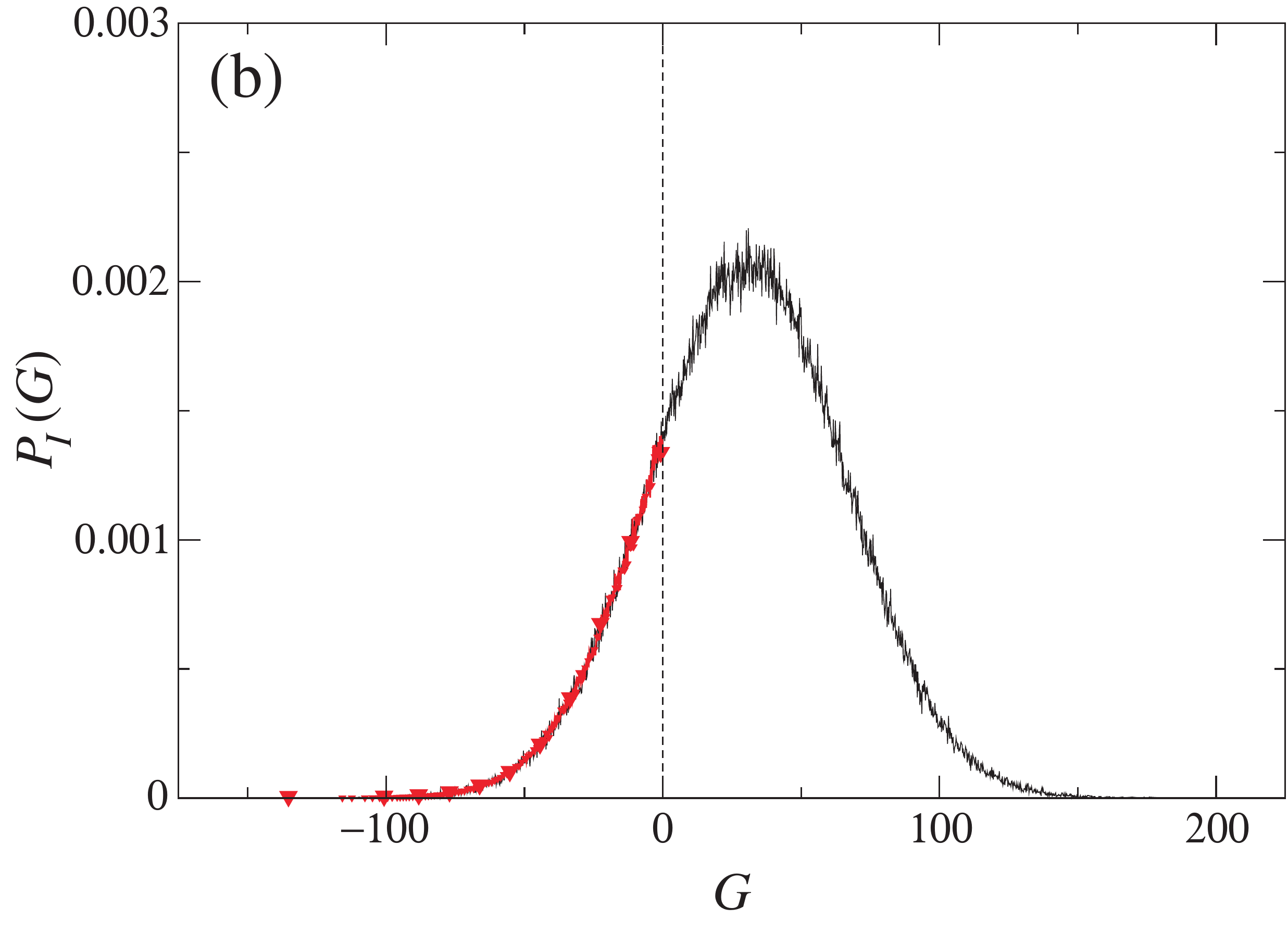}}} \\
\end{tabular}
\caption{(a) Probability distribution of the integrated particle
current, $G=\int_0^t j_k (t') dt'$, evaluated at $k=2$. The
comparison is performed with the fluctuation symmetry \eqref{FT} for
the negative part of the distribution (thick line with triangles).
(b) Probability distribution of the integrated electric current,
$G=(1/e)\int_0^t I (t') dt'$. The comparison is performed with the
fluctuation symmetry \eqref{FTi} for the negative part of the
distribution (thick line with triangles).
The channel is composed of $L=10$ cells, 
each containing $N_- = 25$ fixed ions. The probability
distributions are evaluated at time $t=273$. The channel is
submitted to the boundary conditions $N_{\rm L} = N_{\rm R} = 25$ and
$\phi_{\rm R}-\phi_{\rm L}=0.05$ so that the affinity takes the value
$A=1/20$.}
\label{fig4}
\end{figure}

In this section, we focus on the fluctuations of the particle and
total currents in the zero-frequency limit or, equivalently, in the
long-time limit. More precisely, we will focus on the full
probability distribution of the currents, which contain the
information on the second-order noise properties as well as
information on higher-order properties. In Ref. \cite{AG07JSP} it
was shown in that the probability distribution of the particle
current fluctuations may obey in the long-time limit, $t\rightarrow
\infty$, a fluctuation symmetry of the form
\bea
P_k \left[\frac{1}{t} \int_0^t j_k(t')dt'=\alpha\right] \simeq  P_k 
\left[\frac{1}{t}
\int_0^t j_k(t')dt'=-\alpha\right] {\rm e}^{\alpha A t} \, ,
\label{FT}
\eea
where $A$ is the affinity \eqref{aff}. Here, this symmetry holds as a
result of the thermodynamic properties of our stochastic description,
detailed in Appendix \ref{appB}. An illustration of this relation is
given in Fig. \ref{fig4}a where we depict the probability
distribution function of the particle current, which is compared to the
prediction of the fluctuation theorem \eqref{FT}. 

In addition, we here show that the fluctuation theorem also holds for
the total current \eqref{Itot} exactly in the same form:
\bea
P_I \left[\frac{1}{t} \int_0^t I(t')dt'=\alpha\right] \simeq  P_I
\left[\frac{1}{t} \int_0^t I(t')dt'=-\alpha\right] {\rm e}^{\alpha A
t} \, .
\label{FTi}
\eea
This symmetry can be rigorously shown by extending the demonstration
of Ref. \cite{AG07JSP} to the present situation where any
charge displacement in the system induces a non-local contribution to
the total current, as shown by Eq. \eqref{Itot}. The symmetry
relation \eqref{FTi} is illustrated in Fig. \ref{fig4}b. The fluctuation
relations \eqref{FT} and \eqref{FTi} describe the large nonequilibrium
current fluctuations and have important consequences on the linear and
nonlinear response properties \cite{AG07JSM}.

These results show that the fluctuation theorem is verified for
systems of particles interacting with complex, long-ranged
interaction, whereas most fluctuation relations were verified for
locally interacting particles, for example with purely diffusive 
\cite{AG06JSM} or effusive \cite{CVK06} behavior, or with hard-core interaction \cite{D07}.
For ion transport in channels, the fluctuation theorem was considered 
in the limiting case where the electric repulsion is strong enough to 
prevent the presence of more than one ion inside the channel 
\cite{AG06JSM,BB08,BW08,AG08}.

\section{Conclusions}
\label{conclusions}

In this paper, we have introduced a mesoscopic description of 
ion transport in homogeneous channels 
on the ground of the laws of both electricity and statistical thermodynamics. 
The model is ruled by a master equation describing the
spatial distribution of the discrete numbers of ions in the channel.
Indeed, the stochastic aspects of the time evolution as well as the
discrete nature of the ions are essential aspects of the behavior of
matter at the mesoscopic scale.
The model incorporates the self-consistent field generated by the
fixed and mobile ions as described by the Poisson equation. Also, the
model is shown to be consistent with thermodynamics and can be
studied under equilibrium or nonequilibrium conditions.

The stochastic description can be used to obtain the spatial
correlations of particle density fluctuations in the channel, which reveal the strong repulsion due to
the electric interaction for fluctuations that depart from 
%PG the
electroneutrality. Moreover, we observe long-range correlations in
the electric potential fluctuations.

Another quantity of fundamental interest in the study of ion
transport in channels is the experimentally measured total current, which is composed of the
particle current plus the displacement current arising from the
temporal variations of the electric field. 
At the macroscopic level, the displacement current vanishes in a
stationary state. On the other hand, at the mesoscopic level, the molecular
fluctuations induce long-ranged electric field fluctuations which result in a reduced 
noise spectrum at all positive frequencies. 
The particle
and total currents thus differ in their fluctuation properties and
the present approach allows us to assess the importance of such
effects.
Furthermore, we have shown that the fluctuation theorem for the
currents holds in this more general situation as well, that is when
the displacement current is taken into account in the total current.

In summary, we have shown how to describe ion transport 
in homogeneous channels at the mesoscopic level. 
The extension of the present considerations to the
transport of several ionic species in one channel and to the case of
heterogeneous channels of relevance to biological systems
is possible on the basis of the present results.

\vskip 0.5 cm

{\bf Acknowledgments.} D.~A. is grateful to the F.~R.~S.-FNRS for financial support.
This research is financially supported by the Belgian Federal Government
(IAP project ``NOSY") and the ``Communaut\'e fran\c caise de Belgique''
(contract ``Actions de Recherche Concert\'ees'' No. 04/09-312).

\appendix

\section{Solution of the channel electrostatics}
\label{appA}

The electric potential $\Phi_i$ is defined on each cell $i=1,...,L$
and obeys the discretized Poisson equation \eqref{disc.poisson}
(above written in terms of the dimensionless potential $\phi =
-e\Phi/k_{{\rm B}}T$). This linear system must be solved at each time
the particle distribution changes and must satisfy the boundary
conditions
\bea
\Phi_0=\Phi_{\rm L}, \quad \Phi_{L+1}=\Phi_{\rm R} \, .
\eea

This linear system of equations can be written in matrix form as
\bea
{\pmb{\mathsf C}} \cdot \pmb{\Phi} = {\bf Z}
\eea
with the vectors
\bea
\pmb{\Phi}^{\, {\rm T}} = (\Phi_1, ..., \Phi_L)
\eea
and
\bea
{\bf Z}^{\, {\rm T}} &=& e \paren{N_1-N_-,...,N_L-N_-} + \alpha
(\Phi_{\rm L}, 0, ..., 0, \Phi_{\rm R}) \nonumber \\
&\equiv& {\bf Q} +\alpha (\Phi_{\rm L}, 0, ..., 0, \Phi_{\rm R}) \, ,
\eea
where ${\rm T}$ denotes the transpose operation and where we
introduced the charge vector $\bf Q$ of components $Q_k =
e(N_k-N_-)$. Note that the boundary conditions appear on the first
and last components of the vector $\bf Z$ multiplied by the quantity
$\alpha=(\epsilon N_-)/( n_-\Delta x^2)$, which has the units of a
capacitance.
The $L \times L$ symmetric matrix ${\pmb{\mathsf C}} $ reads
\bea
{\pmb{\mathsf C}} = \alpha
\left(\begin{array}{cccccc}
   2 & -1 &  & & &    \\
   -1 & 2 & -1 &  &  &  \\
    & -1 & 2 & \ddots & & \\
    &  & -1 & \ddots & -1 &  \\
    &  &  & \ddots & 2& -1 \\
    &  &  &  & -1& 2 \\
\end{array} \right)
\eea
It is invertible so that we have
\bea
\pmb{\Phi} = {\pmb{\mathsf C}} ^{-1} \cdot{\bf Z}
\label{Aphi}
\eea
with
\bea
\label{c-1}
\paren{{\pmb{\mathsf C}}^{-1}}_{ij} = \begin{cases}
                   \frac{i}{\alpha (L+1)}(L+1-j) & \text{if} \quad i\leq j \\
                   \frac{j}{\alpha (L+1)}(L+1-i) & \text{if} \quad i > j \, .
                  \end{cases}
\eea
The matrix ${\pmb{\mathsf C}} ^{-1}$ is symmetric, as it should.

The electrostatic energy $U$ stored in the system is given by
\bea
U=\frac{1}{2} \pmb{\Phi}^{\rm T} \cdot  {\pmb{\mathsf C}} \cdot
\pmb{\Phi} = \frac{1}{2} {\bf Z}^{\rm T} \cdot {\pmb{\mathsf C}}^{-1}
\cdot {\bf Z} \, .
\eea
The change in electrostatic energy associated with the transition of
an ion from cell $i$ to cell $i+1$ is thus given by
\bea
\Delta U_{i,i+1} = \frac{1}{2} \paren{{\bf Z'}^{\rm T} \cdot
{\pmb{\mathsf C}} ^{-1} \cdot {\bf Z'} - {\bf Z}^{\rm T} \cdot
{\pmb{\mathsf C}}^{-1} \cdot{\bf Z}}
\eea
where
$Z'_k = Z_k -e\delta_{k,i} +e \delta_{k,i+1}$ characterize the change
in the charge distribution resulting from the transition
$i\rightarrow i+1$.
Developing this expression yields
\bea
\Delta U_{i,i+1} = e (\Phi_{i+1}-\Phi_i) + \frac{e^2}{2}
\Big[\paren{{\pmb{\mathsf C}}^{-1}}_{i,i} -2 \paren{{\pmb{\mathsf
C}}^{-1}}_{i,i+1} +\paren{{\pmb{\mathsf C}}^{-1}}_{i+1,i+1}\Big] \, ,
\label{deii1}
\eea
where we used the symmetry of the coefficients $\paren{{\pmb{\mathsf
C}}^{-1}}_{i,j}=\paren{{\pmb{\mathsf C}}^{-1}}_{j,i}$ as well as the
expression $\Phi_j = \sum_k \paren{{\pmb{\mathsf C}}^{-1}}_{j,k} Z_k$
for the electric potential in cell $j$.
We see that the change in electrostatic energy $\Delta U_{i,i+1}$
only depends on the initial voltage difference plus a term
independent of the charge state of the system.
In the case where the transition involves one of the reservoirs, say the left one,
we have $\Phi'_{\rm L}=\Phi_{\rm L}$ and the change in electrostatic
energy for a transition from the left reservoir to cell $1$ reads
\bea
\Delta U_{0,1} = e (\Phi_{1}-\Phi_{\rm L}) + \frac{e^2}{2}
\paren{{\pmb{\mathsf C}}^{-1}}_{1,1}
\label{de01}
\eea
and similarly for transitions involving the right reservoir.
By virtue of expression \eqref{Aphi}, we see that
\bea
\Phi'_j = \sum_k \paren{{\pmb{\mathsf C}}^{-1}}_{j,k} Z'_k = \sum_k
\paren{{\pmb{\mathsf C}}^{-1}}_{j,k} (Z_k -e\delta_{k,i} +e
\delta_{k,i+1}) = \Phi_j -e \paren{{\pmb{\mathsf C}}^{-1}}_{j,i} +e
\paren{{\pmb{\mathsf C}}^{-1}}_{j,i+1} \, ,
\eea
which is Eq. \eqref{phi'} for the transition $i\rightarrow i+1$.
Accordingly, the energy differences \eqref{deii1} and \eqref{de01}
can be expressed in terms of the initial and final voltage
differences $V_i$ and $V'_i$ between the cells $i$ and $i+1$ so that
we recover Eq. \eqref{dE} for the electrostatic energy difference.
Note that Eq. \eqref{dE} also holds for inhomogeneous channels with
$\epsilon = \epsilon(x)$ and $n_- = n_- (x)$.

\section{Macroscopic affinities of the stochastic model}
\label{appB}

\begin{figure}[htbp]
\centerline{\includegraphics[width=7cm]{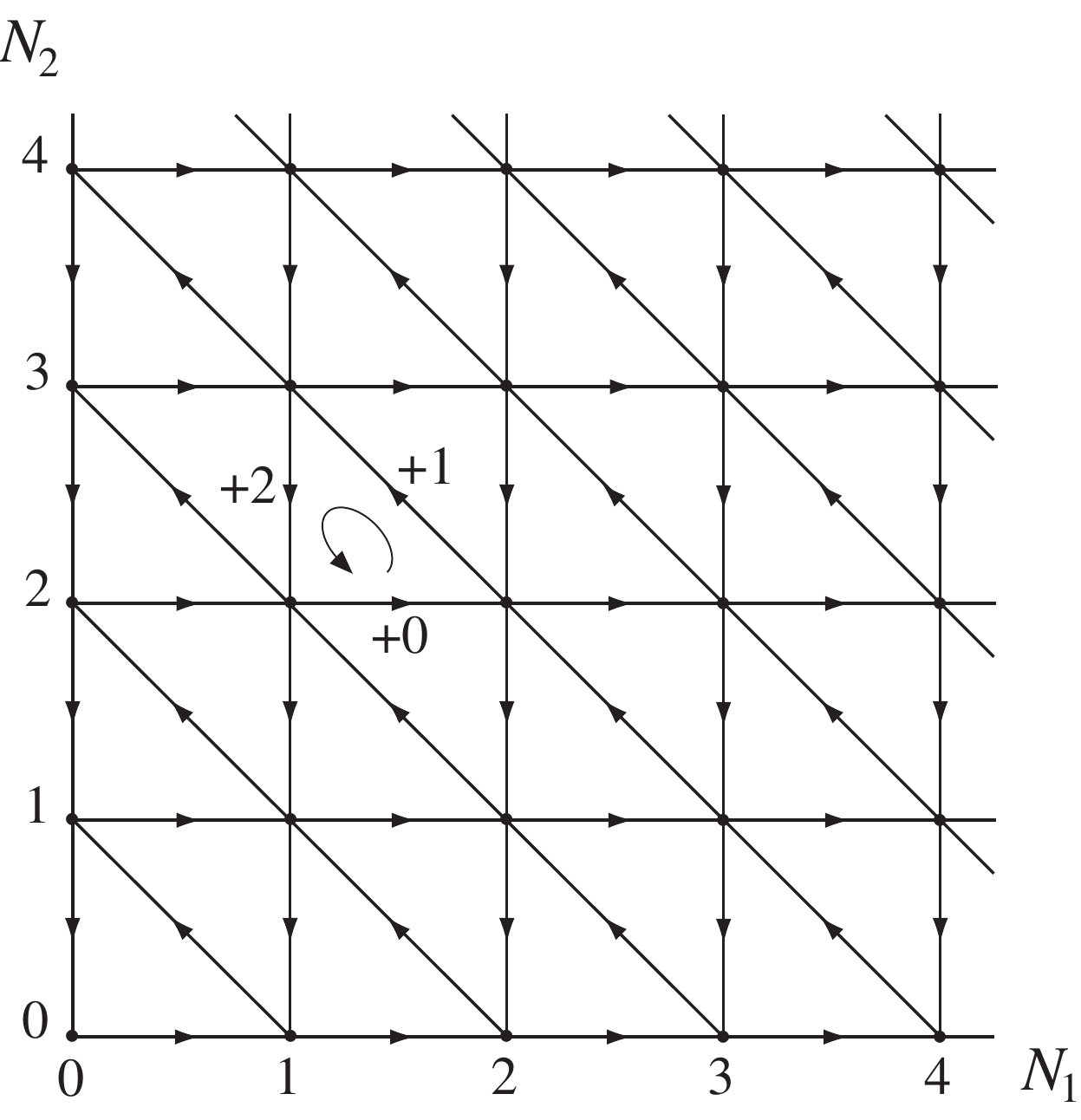}}
\caption{Graph associated with the random process of the master equation
\eqref{ME}
over a chain of length $L=2$. The depicted cycle corresponds to the
transfer of an ion from the left to the right reservoir. The
orientations of the edges are chosen so that a movement from left to
right is counted positively.}
\label{figA1}
\end{figure}

In this appendix, we detail the thermodynamic properties of the
stochastic model introduced in Sec. \ref{stoch}.
For a system ruled by the master equation \eqref{ME},
a graph $G$ is associated as follows \cite{S76}:
each state $\omega$ of the system corresponds to a vertex or node
while the edges represent the different transitions
$\omega \rightleftharpoons \omega'$
allowed between the states.
The stochastic ion transport of Sec. \ref{stoch} is defined in terms
of the populations $(N_1,...,N_L)$ of mobile ions inside the channel.
Accordingly, each such configuration defines a node of the graph and
each transition $\pm i$ corresponds to one edge. As an example the
graph associated with the case $L=2$ is drawn in Fig. \ref{figA1}.

As a matter of fact, cyclic trajectories play a special role as they
link the macroscopic thermodynamic properties to the transition rates
of the mesoscopic description \cite{S76,H05}.
In the case $L=2$ corresponding to Fig. \ref{figA1}, we see that any
cyclic path in the graph can be decomposed as a linear combination of
two types of cyclic trajectories:
\bea
c_1 \equiv (N_1, N_2) \rightarrow  (N_1+1, N_2) \rightarrow  (N_1+1,
N_2+1) \rightarrow  (N_1, N_2+1) \rightarrow  (N_1, N_2)
\eea
and
\bea
c_2 \equiv (N_1, N_2) \rightarrow  (N_1+1, N_2) \rightarrow  (N_1,
N_2+1) \rightarrow  (N_1, N_2) \, .
\eea
As shown by Schnakenberg \cite{S76}, the macroscopic affinities
associated with these cyclic paths can be obtained from the
transition rates by calculating the quantities
\bea
A(c_1) = \ln \frac{W_{+0}(N_1,N_2) W_{-2}(N_1+1,N_2) W_{-0}(N_1+1,N_2
+1) W_{+2}(N_1,N_2 +1) }{W_{-0}(N_1+1,N_2) W_{+2}(N_1+1,N_2 +1)
W_{+0}(N_1,N_2 +1) W_{-2}(N_1,N_2)}
\label{Ac1}
\eea
and
\bea
A(c_2) = \ln \frac{W_{+0}(N_1,N_2) W_{+1}(N_1+1,N_2)
W_{+2}(N_1,N_2+1)}{W_{-0}(N_1+1,N_2) W_{-1}(N_1,N_2+1)
W_{-2}(N_1,N_2)} \, .
\label{Ac2}
\eea
These expressions are obtained as the ratio between the product of
the transition rates along the cyclic path in one direction over the
product of the transition rates along the cyclic path in the reversed
direction.
Calculating the affinities \eqref{Ac1} and \eqref{Ac2} with the
transition rates \eqref{W} yields
\bea
A(c_1) = 0 \quad {\rm and} \quad A(c_2) = \ln
\paren{\frac{N_0}{N_{L+1}}{\rm e}^{\Delta\phi}} \, .
\label{ac12}
\eea
The first one always vanishes whereas the second is the macroscopic
affinity \eqref{aff}, irrespectively of the initial configuration
$(N_1,...,N_L)$. Note that $A(c_2)$ only involves externally
controlled parameters, as it should. This results from the fact that
only the second cycle involves the transport of an ion across the
channel. The extension to larger channels, $L > 2$, is
straightforward.

According to a result of Kolmogorov \cite{K36}, a Markov process is
at equilibrium if and only if the affinities along the cycles vanish.
As shown by Eqs. \eqref{ac12} this occurs when condition \eqref{eq}
is fulfilled.

%%%%%%%%%%%%%%%%%%%%%%%%%%%%%%%%%%%%%%%%%%%%%%%%%%%%%%%%%%%%%%%%%%%%%%%%%%%%%%%%%%%%%%%%%%%%%%%%%%%%%%


\begin{thebibliography}{99}
\bibitem{footnote} Several functions $\psi(\Delta U)$ can be chosen which satisfy
the condition (\ref{det_bal}).  Besides the function (\ref{dfn_psi})
which has been considered in another context \cite{WPE03}, alternative functions
are the Metropolis one $\psi(\Delta U)=(D/\Delta x^2){\rm min}\left\{1,{\rm e}^{-\beta\Delta U}\right\}$,
or the Kawasaki one $\psi(\Delta U)=(D/\Delta x^2) [2 /\left(1+{\rm e}^{\beta\Delta U}\right)]$, 
which also satisfy the condition (\ref{det_bal}) \cite{Spohn}. 
The specific form of the function $\psi(\Delta U)$
should be determined by the microscopic process ruling the jumps of particles between the cells.
We emphasize that, in our case, the energy difference $\Delta U$ depends
on the global configuration because of the long-range Coulomb interaction.  
This dependence is required in order to obtain a model 
which is thermodynamically consistent, irrespectively of the specific choice of the function $\psi$.
\end{thebibliography}
\end{document}